\documentclass[lettersize,journal]{IEEEtran}

\usepackage{amsmath,amsfonts,amsthm,amssymb}
\usepackage{algorithmic}
\usepackage{algorithm}
\usepackage{array}
\usepackage{textcomp}
\usepackage{stfloats}
\usepackage{url}
\usepackage{xcolor,soul,framed} 
\usepackage{xcolor}
\usepackage{verbatim}
\usepackage{graphicx}
\usepackage{subcaption}
\usepackage{epstopdf}
\usepackage{cite}
\begin{document}
\pdfoutput=1
\newtheorem{Remark}{Remark}
\newtheorem{remark}{Remark}
\renewcommand{\algorithmicrequire}{\textbf{Input:}} 
\renewcommand{\algorithmicensure}{\textbf{Output:}}

\newcommand{\zhenyu}[1]{{\color{blue}\sf{[Zhenyu: #1]}}}

\title{RIS-aided Wireless Communication with $1$-bit Discrete Optimization for Signal Enhancement}

\author{Rujing~Xiong,~\IEEEmembership{Student Member,~IEEE,}
        Xuehui~Dong,~\IEEEmembership{Student Member,~IEEE,}
        Tiebin~Mi,~\IEEEmembership{Member,~IEEE,}
 		Robert~Caiming~Qiu,~\IEEEmembership{Fellow,~IEEE,}
\thanks{R.~Xiong, X.~Dong, T.~Mi, and R.~Qiu are with the School of Electronic Information and Communications, Huazhong University of Science and Technology, Wuhan 430074, China (e-mail: \{rujing,~dong\_xh,~mitiebin,~caiming\}@hust.edu.cn).}
\thanks{National Foundation, NO.12141107, supports this work.}
\thanks{Manuscript received xxx, 2022; revised xxx, 2022.}
}

\markboth{Journal of \LaTeX\ Class Files, xxx, xxx, xxx, 2022}
{Shell \MakeLowercase{\textit{et al.}}: A Sample Article Using IEEEtran.cls for IEEE Journals}


\maketitle

\begin{abstract}
In recent years, a brand-new technology, reconfigurable intelligent surface (RIS) has been widely studied for reconfiguring the wireless propagation environment. RIS is an artificial surface of electromagnetic material that is capable of customizing the propagation of the wave impinging upon it. Utilizing RIS for communication service like signal enhancement usually lead to non-convex optimization problems. 
Existing optimization methods either suffers from scalability issues for $N$ number of RIS elements large, or may lead to suboptimal solutions in some scenario. 
In this paper, we propose a divide-and-sort (DaS) discrete optimization approach, that is guaranteed to find the global optimal phase shifts for $1$-bit RIS, and has time complexity  $\mathcal{O}(N \log(N))$. 
Numerical experiments show that the proposed approach achieves a better ``performance--complexity tradeoff'' over other methods for $1$-bit RIS.
\end{abstract}


\section{Introduction}

\IEEEPARstart{T}{he} reconfigurable intelligent surface (RIS) technique has recently demonstrated its great potential for reconfiguring the wireless propagation environment via software-programmed reflection~\cite{cui2014coding,basar2019wireless,di2020smart}. 
A RIS consists of a large number of carefully designed electromagnetic cells and can 
result in electromagnetic fields with controllable behaviors such as amplitude, phase, polarization, and frequency. It revolutionizes the traditional communication environment that cannot be controlled and leads to a change of paradigm in wireless communication.
RISs can be implemented in wireless communication networks to improve performance, e.g., signal-to-noise ratio (SNR).



To apply RIS for signal enhancement in wireless communication, one needs to design a phase shift matrix that has unit module diagonal entries by, maximizing the SNR.
Such optimization problem has non-convex constraints and is NP-hard in general~\cite{wu2019intelligent}. Considerable effort has been made to solve these problems in an accurate and efficient manner: 
(i) semi-definite relaxation (SDR) was used to relax the non-convex constraints, such that the transformed optimization problem can be efficiently solved by semi-definite programming (SDP)~\cite{cui2019secure,elmossallamy2021ris,zhou2020robust};  
(ii) manifold optimization (Manopt) methods~\cite{absil2009optimization,yu2019miso, elmossallamy2021ris,hu2022constant}, utilize optimization methods such as gradient descent on the complex circle manifold imposed by the constraints. 
These two approaches were designed for continuous phase shift matrices, where the obtained phase shifts can take \emph{any} value within $[-\frac{\pi}2,\frac{3\pi}2)$. 
 For low-bits RIS, a natural discrete approach is to perform exhaustive search for the phase shifts matrix. This always leads to an optimal solution but with a time complexity exponential with the number of RIS elements $N$. 
More recently, discrete design for RIS phase shifts has attracted a unprecedentedly research interest.
In \cite{wu2019beamforming} the authors proposed to find the (discrete) phase shifts matrix via discrete optimization, this significantly accelerates the optimization, while the obtained solution may be sub-optimal.
\cite{zhang2022configuring} introduced an approximation (APX) algorithm to solve the discrete optimization problem, which achieves the global optimality when the number of states for each component of the RIS is large enough.

1-bit RIS is the most common setting in the existing prototype realizations, e.g.,~\cite{chen2020active, arun2020rfocus, tran2020demonstration, pei2021ris, kitayama2021research, staat2022irshield}. It is hard to obtain the global optimal phase shifts due to the integer programming. To the best of our knowledge, there is no existing can achieve optimality with a linear or near linear time complexity.
In this paper, we propose a divide-and-sort (DaS) approach to the discrete optimization problem that (i) has time complexity $\mathcal{O}(N \log(N))$
; and (ii) is guaranteed to find the global optimal solution. 

\medskip
\emph{Notations}: Lower and upper case bold fonts denote vectors(e.g., $\mathbf{a}$) and matrix (e.g., $\mathbf{A}$).  $\mathbf{A}^H$, $\mathbf{A}^*$,  and $\mathbf{A}^T$ denotes the conjugate transpose, conjugate, and transpose of $\mathbf{A}$, respectively. $\mathbb{C}^{a\times b}$ denotes the field of complex-valued matrices of dimension $a\times b$. 
$\Vert \mathbf{a}\Vert$ represents the Frobenius norm of $\mathbf{a}$ and $\rm {diag}(\mathbf{a})$ is a diagonal matrix with each diagonal element being the corresponding element in $\mathbf{a}$. For a complex number $a \in \mathbb{C}$, we use $j$ as the imaginary unit and denote its modulus, real part, and angle as $|a|$, $\mathcal{\Re} \{a\}$ and $\rm {arg}\{ a\}$, respectively. $\rm {sgn}\{a\}$ is the sign of real number $a \in \mathbb{R}$.  

\medskip
The remainder of the paper is organized as follows. Section \ref{Section2} introduces the system model and problem formulation for signal enhancement in the RIS-aided communication system. In Section \ref{Section3}, we present the DaS discrete optimization method to solve the problem. Section \ref{Section4} presents numerical results to evaluate the performance and time efficiency of the proposed algorithm with respect to some other such as APX, Manopt, and SDR-SDP approaches. Finally, we conclude the paper in Section \ref{Section5}.

\section{System model and problem formulation}\label{Section2}
\subsection{System Model}
\begin{figure}
\centering
\begin{subfigure}{.24\textwidth }
\centering
\includegraphics[width=.9\linewidth]{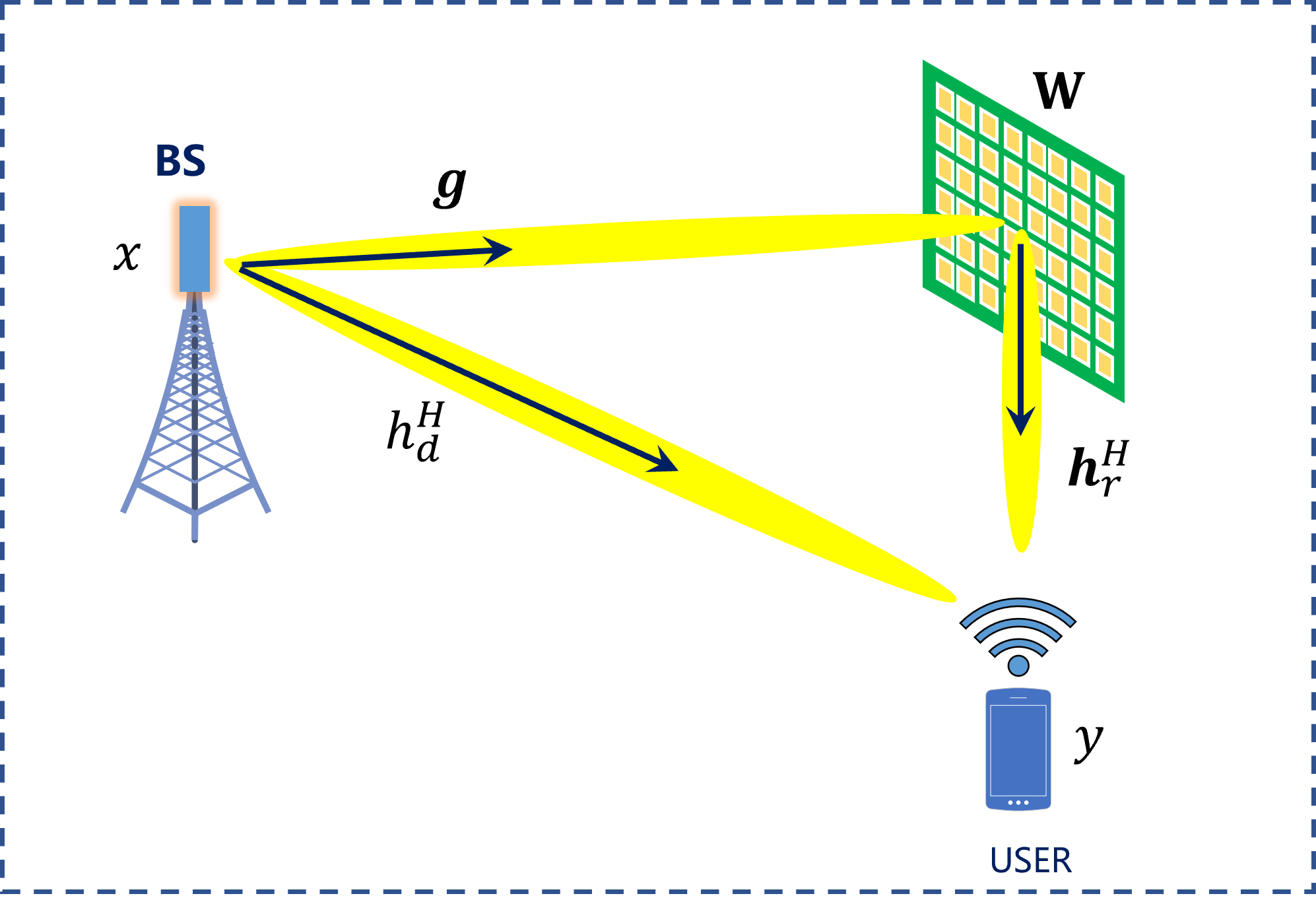}%
\caption{with direct link}
\label{With LoS}
\end{subfigure}
\hfil
\begin{subfigure}{.24\textwidth}
\centering
\includegraphics[width=.9\linewidth]{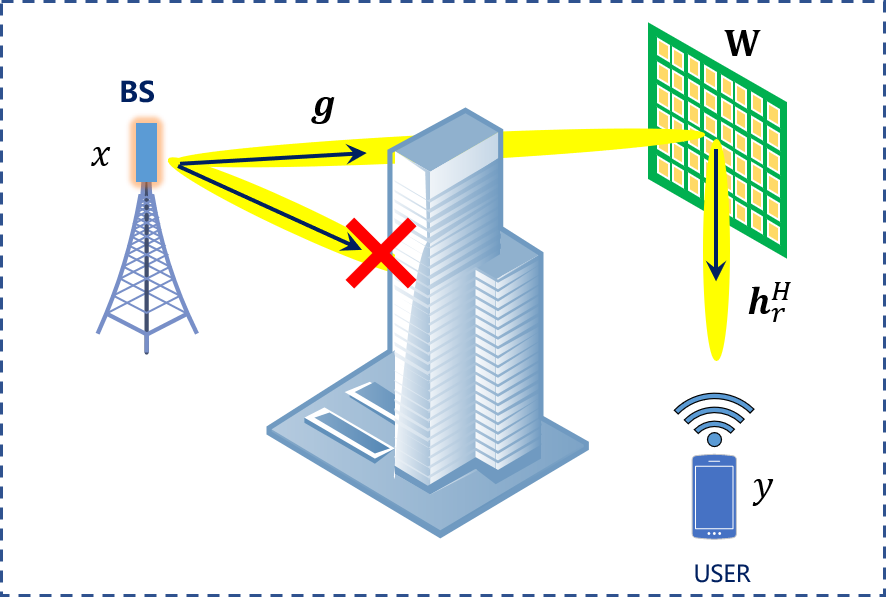}%
\caption{without direct link}
\label{With NLoS}
\end{subfigure}
\caption{A RIS composed of $N$ passive elements employed for a single input single output communication system service.}
\label{Scenarios}
\end{figure}

We consider a single input single output communication scenario as shown in Fig.\ref{With LoS}. The user receives the signals from a single antenna transmitter, and a RIS composed of $N$ passive elements is employed to improve the communication quality of the user. Due to substantial path loss, it is assumed that the power of signals reflected by the RIS two or more times is negligible and thus ignored. There is a direct link between the transmitter and receiver, and the user can receive signals from both the direct and reflected links. Then, the total received signals from the base station (BS) can be expressed as\cite{wu2019intelligent}: 
\begin{equation}
y = (\mathbf{h}_r^H\mathbf{W}(\boldsymbol{ \theta })\mathbf{g}+h_d^H)x +s,
\end{equation}
where $y \in \mathbb{C}$ is the received signal, $x \in \mathbb{C}$ is the modulated signal from the BS, and $s \in \mathbb{C}$ denotes the additive Gaussian white noise. The equivalent channel between BS-RIS, RIS-USER, and RIS-USER links can be represented by $\mathbf{g} \in \mathbb{C}^{N\times 1}, \mathbf{h}_r^H\in \mathbb{C}^{1\times N}$ and $h_d^H \in \mathbb{C} $,  respectively. 
The passive RIS changes the phase of the incident signals, let $\theta_n \in \left[-\frac{\pi}{2},\frac{3\pi}{2}\right)$ denote the phase shift associated with the $n$-th passive element of the RIS. Define $\boldsymbol{ \theta } = \left[\theta_1,\theta_2,\dots,\theta_N\right] \in \mathbb{R}^N$, and $\mathbf{W}(\boldsymbol{ \theta }) = {\rm diag} (w_1,w_2,\dots,w_N)\in \mathbb{C}^{N\times N}$ as the reflection-coefficients matrix of the RIS, where $w_n=e^{j\theta_n}, n=1,...,N$. 

\subsection{Problem Formulation}
Our goal is to maximize the received signal power, by solving the  following optimization problem:

\begin{equation}\label{P1}
\begin{aligned}
\rm
\mathop{max}\limits_{\emph{x},\boldsymbol{ \theta }}& \ \left |(\mathbf{h}_r^H\mathbf{W}(\boldsymbol{ \theta })\mathbf{g}+h_d^H)x\right|^{2},\\
s.t.& \ -\frac{\pi}{2}\leq \theta_n<\frac{3\pi}{2},\quad \forall n=1, \cdots, N. 
\end{aligned}
\end{equation}

For any given phase shift vector $\boldsymbol \theta \in \mathbb{R}^N$, it can be verified that the maximum-ratio transmission (MRT) is the optimal transmission solution\cite{tse2005fundamentals}. The problem can thus be simplified as the following equivalent problem:

\begin{equation}\label{P2}
\begin{aligned}
\rm
\mathop{max}\limits_{\boldsymbol{ \theta }}& \ \left\|(\mathbf{h}_r^H\mathbf{W}(\boldsymbol{ \theta })\mathbf{g}+h_d^H)\right\|^{2},\\
s.t.& \ \left|w_n\right|=1, \quad\forall n=1, \cdots,N.
\end{aligned}
\end{equation}

In a $1$-bit RIS-aided wireless communication system, the phase shifts $\theta_n$ caused by each RIS component on the incoming wave take one of the possible two values with a difference of $\pi$. Here, we consider one of the state values to be 0 and the other to $\pi$ as in \cite{cui2014coding,tan2018enabling,wu2019beamforming}.
In this case, the optimization problem in (\ref{P2}) can be formulated as:
\begin{equation}\label{PP}
\begin{aligned}
\rm
\mathop{max}\limits_{\boldsymbol{ \theta }}& \ \left\|(\mathbf{h}_r^H\mathbf{W}(\boldsymbol{ \theta })\mathbf{g}+h_d^H)\right\|^{2},\\
s.t.& \ \mathbf{ w}\in \{\pm 1\}^{N},
\end{aligned}
\end{equation}
where $\mathbf{w} =[w_1,w_2,\dots,w_N]^T\in \mathbb{R}^N$.
Denote  the shortcut $\boldsymbol{ \phi } = {\rm diag}(\mathbf{h}_r^H)\mathbf{g} \in \mathbb{C}^{N}$. The problem can be further written as:
\begin{equation}\label{P3}
\begin{aligned}
\rm
\mathop{max}\limits_{\mathbf{w}}& \ \mathbf{w}^T\boldsymbol{ \phi } \boldsymbol{ \phi }^H \mathbf{w}+h_d^H \boldsymbol{ \phi }^H \mathbf{w}+\mathbf{w}^T \boldsymbol{ \phi } h_d+\left| h_d\right|^2,\\
s.t.& \ \mathbf{ w}\in \{\pm 1\}^{N}.
\end{aligned}
\end{equation}

The problem in (\ref{P3}) is a non-convex quadratically constrained quadratic program \cite{so2007approximating}, which, by introducing a constant $1$, can be turned into the following standard homogeneous quadratic programming problem:

\begin{equation}\label{P5}
\begin{aligned}
\rm
\mathop{max}\limits_{\mathbf{\bar w}}& \ \mathbf{\bar w}^T\mathbf{R}\mathbf{\bar w}, \\
s.t.& \ \mathbf{\bar w}\in \{\pm 1\}^{N+1}.
\end{aligned}
\end{equation}

where $\mathbf{\bar{w}}=\left[\mathbf{w}^T, 1 \right]^T \in \mathbb{R}^{N+1}$, $\boldsymbol{\bar{ \phi }} = \left[\begin{array}{c}\boldsymbol{ \phi }^T, h_d^H \end{array}\right]^T  \in \mathbb{C}^{{N+1}}$. Note that $\mathbf{R}= \boldsymbol{\bar{ \phi }}\boldsymbol{\bar{ \phi }}^H$, and is thus a rank-one and semi-positive definite Hermitian matrix.

\begin{Remark}[Special case: without line of sight]
The direct link (or line of sight, LoS) between the BS and the user may, in some scenarios, blocked by obstacles such as buildings, as shown in Fig. \ref{With NLoS}. In this case, $h_d^H = 0$ in (\ref{P2}), and the optimization problem to achieve the maximum received signal power can be further reduced as:
  \begin{equation}\label{without los}
  \begin{aligned}
  \rm
  \mathop{max}\limits_{\mathbf{w}}& \ \mathbf{w}^T\mathbf{R}\mathbf{w},\\
  s.t.& \ \mathbf{w}\in \{\pm 1\}^N,
  \end{aligned}
  \end{equation}
where $\mathbf{R} = \boldsymbol{ \phi } \boldsymbol{ \phi }^H$, again of rank $1$ and semi-positive definite.
\end{Remark}

\section{$1$-bit Phase Shift Optimization}\label{Section3}
It is an integer optimization problem in (\ref{P5}). A naive, but computationally less efficient approach is to perform greedy search among all elements of $\{\pm 1\}^{N+1}$, which has time complexity $\mathcal{O}(2^{N+1})$. In this section, we propose a novel and efficient algorithm that significantly reduces the time complexity from $\mathcal{O}(2^{N+1})$ to $\mathcal{O}(N\log{N})$.
Our proposed algorithm mainly contains the following steps: (i) compute the phase of the eigenvector of $\mathbf{R}$; (ii) introduce an auxiliary variable and take the divide-and-sort strategy to construct the set $\mathcal{U}$, which contains the optimal solution; (iii) perform an exhaustive search for the optimal solution among $\mathcal{U}$.

Since the Hermitian matrix $\mathbf{R}\in \mathbb{C}^{(N+1)\times(N+1)}$ is of rank one, it can be spectrally decomposed as:  
\begin{equation}
\mathbf{R} = \lambda \mathbf{z} \mathbf{z}^H,
\end{equation}
where the $\lambda>0$ is the (only non-zero) eigenvalue, and $\mathbf{z} \in \mathbb{C}^{N+1}$ corresponding eigenvector of $\mathbf{R}$.\\
Thus, the problem in (\ref{P5}) becomes:

\begin{equation}
\begin{aligned}
\rm
\mathop{max}\limits_{\mathbf{\bar w}}& \ \mathbf{\bar w}^T\lambda \mathbf{z} \mathbf{z}^H\mathbf{\bar w},\\
s.t. &\ \mathbf{\bar w}\in \{\pm 1\}^{N+1}.
\end{aligned}
\end{equation}
The problem further writes

\begin{equation}\label{$1$-bit}
\begin{aligned}
\rm
\mathop{max}\limits_{\mathbf{\bar w}}& \ \left|\mathbf{\bar w}^T\mathbf{z}\right|,\\
s.t.& \ \mathbf{\bar w}\in \{\pm 1\}^{N+1}.
\end{aligned}
\end{equation}
Recall 
 $\mathbf{\bar w}=[w_1,w_2,\dots,w_{N+1}]^T \in \mathbb{R}^{N+1}$, and let $\mathbf{z}=[z_1,z_2,\dots,z_{N+1}]^T \in \mathbb{C}^{N+1}$ with polar decomposition:
\begin{equation}
z_n = |z_n|e^{j\theta_n}, -\frac{\pi}{2}\leq \theta_n<\frac{3\pi}{2},\quad n = 1,\dots, {N+1}.
\end{equation}
Following the idea in \cite{karystinos2007rank}, we introduce an auxiliary variable~$\psi \in[-\frac{\pi}{2},\frac{3\pi}{2})$, and rewrite the quantity to be maximized in (\ref{$1$-bit}) as:
\begin{small}
\begin{equation}
\begin{aligned}
|\mathbf{\bar w}^T\mathbf{z}| = \rm
		\mathop{max}\limits_{\psi \in[-\frac{\pi}{2},\frac{3\pi}{2})}& \
        \Re\{\mathbf{\bar w}^T\mathbf{z}e^{-j\psi}\}\\
       =\rm
		\mathop{max}\limits_{\psi \in[-\frac{\pi}{2},\frac{3\pi}{2})}& \ \left\{ \sum_{n = 1}^{N+1} \bar{w}_n|z_n|\cos(\psi-\theta_n)   \right\}.
\end{aligned}
\end{equation}
Then,
\begin{subequations}
\begin{align}
& {\rm \mathop{max}\limits_{\mathbf{\bar w}\in \{\pm 1\}^{{\emph N}+1}}} \ \left|\mathbf{\bar w}^T\mathbf{z}\right|  \notag \\
        = &{\rm
          \mathop{max}\limits_{\mathbf{\bar w}\in \{\pm 1\}^{{\emph N}+1}}} {\rm \mathop{max}\limits_{\psi \in[-\frac{\pi}{2},\frac{3\pi}{2})}} \  \;\left\{ \sum_{n = 1}^{N+1} \bar{w}_n|z_n|\cdot \cos(\psi-\theta_n)\right\}   \label{Za}\\
        = &{\rm
		\mathop{max}\limits_{\psi \in[-\frac{\pi}{2},\frac{3\pi}{2})} }\  \left\{ \sum_{n = 1}^{N+1}\left|z_n\right|\cdot \left|\cos(\psi-\theta_n)\right| \right\}  \label{Zb}
\end{align}
\end{subequations}
\end{small}
where we used in the second equality the fact that for any $\psi \in[-\frac{\pi}{2},\frac{3\pi}{2})$ the optimal $\bar{w}_{n}$ in (\ref{Za}) is simply $\bar{w}_{n}(\psi)=\operatorname{sgn}\left(\cos \left(\psi-\theta_{n}\right)\right), n=1,2, \ldots, N+1$. The final quantity $\sum_{n=1}^{N+1}\left|z_{n}\right| \left|\cos \left(\psi-\theta_{n}\right)\right|$ in (\ref{Zb}) is maximized for a particular value $\psi_{opt}\in[-\frac{\pi}{2},\frac{3\pi}{2})$ and $\bar{\mathbf{w}}\left(\psi_{opt}\right) = [w_1(\psi_{opt}), w_2(\psi_{opt}),\dots,w_{N+1}(\psi_{opt})]^T$ is the optimal binary vector we are searching for in (\ref{$1$-bit}), i.e. $\bar{\mathbf{w}}_{2}=\bar{\mathbf{w}}\left(\psi_{opt}\right)$. We will now show that we can always construct a set of $N+1$ binary spreading codes ${\mathcal{U}}=\left\{\mathbf{u}_{1}, \mathbf{u}_{2}, \ldots, \mathbf{u}_{N+1}\right\}, \mathbf{u}_{n} \in\left\{\pm 1\right\}^{N+1}$, and guarantee that $\bar{\mathbf{w}}\left(\psi_{opt}\right) \in {\mathcal{U}}$. Therefore, the maximization in (\ref{$1$-bit}) can be performed on the set ${\mathcal{U}}$ of $N+1$ candidates only without loss of optimality.

To construct the set ${\mathcal{U}}$, we partition the index set $\mathcal{Z}_{N+1}=\{1,\cdots, {N+1}\}$ into 
\begin{equation}
\begin{aligned}
&I_{1} \triangleq \left\{n:  \theta_{n} \in \left[-\frac{\pi}{2}, \frac{\pi}{2}\right)\right\},\\
&I_{2} \triangleq \left\{n:  \theta_{n} \in \left[\frac{\pi}{2}, \frac{3 \pi}{2}\right)\right\}=\mathcal{Z}_{{N+1}}\setminus I_{1}
\end{aligned}
\end{equation}
and define the angles
\begin{equation}\label{angles}
\begin{aligned}
\hat{\theta}_{n} \triangleq\left\{\begin{array}{cc}
\theta_{n}, & n \in I_{1} \\
\theta_{n}-\pi, & n \in I_{2}
\end{array}\quad, \quad n=1, \cdots, {N+1},\right.
\end{aligned}
\end{equation}
so that $\hat{\theta}_{n}\in \left[-\frac{\pi}{2},\frac{\pi}{2}\right)$, for all $n=1, \cdots, {N+1}$. Further define, for notational simplicity, the vector operation $\hat{\mathbf{w}}(\psi) \triangleq\left[\hat{w}_{1}(\psi) \quad \hat{w}_{2}(\psi) \quad \cdots \quad \hat{w}_{N+1}(\psi)\right]^{T} $ with $ \hat{w}_{n}(\psi) \triangleq$ $\operatorname{sgn}\left(\cos \left(\psi-\hat{\theta}_{n}\right)\right), n=1, \cdots, {N+1}$. Then, one has
\begin{small}
\begin{equation}
\begin{aligned}
w_{n}(\psi)=\left\{\begin{array}{rl}
\hat{w}_{n}(\psi), & n \in I_{1} \\
-\hat{w}_{n}(\psi), & n \in I_{2}
\end{array} \quad, \quad n=1,\cdots, {N+1}.\right.
\end{aligned}
\end{equation}
\end{small}

Consider a mapping $m$ from $\mathbb{R}^{N+1}$ to $\mathbb{R}^{N+1}$ that sorts the angles $\hat{\theta}_{1},\hat{\theta}_{2},\cdots, \hat{\theta}_{N+1}$ in a non-decreasing order $-\frac{\pi}{2}\leq \hat{\theta}_{m(1)}\leq \hat{\theta}_{m(2)}\leq \dots \leq \hat{\theta}_{m({N+1})}<\frac{\pi}{2}$. It follows from \eqref{Za}, \eqref{Zb}, that the optimal binary vector $\hat{\mathbf{w}}(\psi)=\left[\begin{array}{lllll}
\hat{w}_{m(1)}(\psi) & \hat{w}_{m(2)}(\psi) & \cdots & \hat{w}_{m({N+1})}(\psi)
\end{array}\right]$ we are searching for can be obtained from the construct $\tilde {\mathcal{U}}$ as bellow\cite{karystinos2007rank}:
\begin{footnotesize}
\begin{equation}\label{setU}
\begin{aligned}
& \tilde{\mathcal{U}}= 
&\begin{array}{cl}
 \left\{ \underbrace{+1, \cdots,+1,}_{n} \underbrace{-1, \cdots,-1}_{{N+1}-n}\right\}, \quad \hat{\theta}_{m(n)}-\frac{\pi}{2} \leq \psi<\hat{\theta}_{m(n+1)}-\frac{\pi}{2}\\
\end{array},
\end{aligned}
\end{equation}
\end{footnotesize}
where $ n=1, \cdots, {N+1}$, note that while $n = N+1$, $n+1$ returns back to 1. We collect the ${N+1}$ binary vectors in the ${N+1}$ cases $\tilde{\mathbf{u}}_{n} \triangleq[\underbrace{+1, \cdots,+1,}_{n} \underbrace{-1, \cdots,-1}_{{N+1}-n}]^T, n=1,\cdots, {N+1}$  as the matrix 
\begin{equation} \label{tildeU}
\tilde {\mathbf{U}}\triangleq [\tilde{\mathbf{u}}_{1},\tilde{\mathbf{u}}_{2},\cdots,\tilde{\mathbf{u}}_{N+1}].
\end{equation}
Then, we reorganize $ \tilde{\mathbf{U}}$ to $\hat{\mathbf{U}}\triangleq \left[\begin{array}{llll}\hat{\mathbf{u}}_{1}, & \hat{\mathbf{u}}_{2}, & \ldots, & \hat{\mathbf{u}}_{N+1}\end{array}\right]$ by defining the binary vectors
\begin{equation}
\hat{\mathbf{u}}_{n} \triangleq \tilde{\mathbf{u}}_{m^{-1}(n)}, \quad n=1,2, \cdots, N+1,
\end{equation}
where $m^{-1}: \mathbb{R}^{N+1} \rightarrow \mathbb{R}^{N+1}$ is the inverse sorting mapping, note that $\hat{\mathbf{w}}(\psi) \in\left\{\hat{\mathbf{u}}_{1}, \hat{\mathbf{u}}_{2}, \ldots, \hat{\mathbf{u}}_{N+1}\right\}$ for any $\psi \in$ $\left.\left[\hat{\psi}_{m(1)}-\frac{\pi}{2},   \hat{\psi}_{m(1)}+\frac{\pi}{2}\right)\right.$ Finally, we take
\begin{equation}
\mathbf{u}_{n} \triangleq\left\{\begin{array}{cc}
 \hat{\mathbf{u}}_{n}, & n \in I_{1} \\
 -\hat{\mathbf{u}}_{n}, & n \in I_{2}
\end{array} \quad, \quad n=1,2, \cdots, N+1,\right.
\end{equation}
and construct
\begin{equation}\label{U}
\mathbf{U}=\left[\begin{array}{llll}
\mathbf{u}_{1}, & \mathbf{u}_{2}, & \ldots, & \mathbf{u}_{N+1}
\end{array}\right].
\end{equation}
The set $\mathcal{U} \triangleq\left\{\mathbf{u}_{1}, \mathbf{u}_{2}, \cdots, \mathbf{u}_{N+1}\right\}$ contains $\bar{\mathbf{w}}\left(\psi\right)$ for any $\psi \in$ $\left[\hat{\psi}_{m(1)}-\frac{\pi}{2}, \hat{\psi}_{m(1)}+\frac{\pi}{2}\right)$. And $\psi_{opt} \in\left[\hat{\psi}_{m(1)}-\frac{\pi}{2}, \hat{\psi}_{m(1)}+\frac{\pi}{2}\right)$, which implies $\bar{\mathbf{w}}\left(\psi_{opt}\right) \in \mathcal{U}$.
Hence, the optimization solution to (\ref{P5}) becomes:
\begin{equation}\label{discrete optimization}
\bar{\mathbf{w}}\left(\psi_{opt}\right) = {\rm \arg} \ {\rm \mathop{max}\limits_{\mathbf{\bar w}\in \mathcal{U}}}\ \left|\mathbf{\bar w}^T\mathbf{z}\right| \triangleq \mathbf{ \bar{w}}_{opt}.
\end{equation}

 Finally, the solution to problem (\ref{PP}) can be obtained by $\mathbf{ w}_{opt} = (\mathbf{ \bar{w}}_{opt}^{(1:N)}./{ \bar{w}}_{opt}^{(N+1)})$, where $\left[ \mathbf{x}\right]^{(1:N)}$ and$\left[ {x}\right]^{(N+1)}$ denotes the vector that contains the first $N$ elements and the $(N-1)$-th element in $\mathbf{x}$, respectively, operator $'./'$ here is element-wise (right) scalar division, which returns the vector that elements are divided by the scalar. Phase shifts matrix is $\mathbf{W}(\boldsymbol{ \theta })={\rm{diag}}(\mathbf{ w}_{opt}^H)$. The complexity of the construction of $\mathbf{U}$ is dominated by the complexity of the mapping function $m$, which is of order $\mathcal{O}({N}\log{{N}})$. 

The proposed algorithm is summarized in Algorithm~\ref{alg2}.

\begin{algorithm}
\caption{Divide-and-sort (DaS) phase shifts optimization for $1$-bit RIS } 
\label{alg2} 
\begin{algorithmic}[1] 
\REQUIRE Equivalent channel matrices $\mathbf{g}, \mathbf{h}_r^H$, and ${h}_d^H$ 
\ENSURE Phase shifts matrix $\mathbf{W}(\boldsymbol{ \theta })$ 
\STATE Calculate matrix $\mathbf{R}=\boldsymbol{ \bar{\phi} }\boldsymbol{ \bar{\phi} }^H = \left[{\rm diag}(\mathbf{h}_r^*)\mathbf{g}^T, {h}_d^H \right]^T \cdot    \left[\mathbf{g}^H{\rm diag}(\mathbf{h}_r), {h}_d \right]$ from $\mathbf{g}, \mathbf{h}_r^H$, and $h_d^H$
\STATE Perform eigenvalue decomposition $\mathbf{R}=\lambda \mathbf{z} \mathbf{z}^H$ to obtain the eigenvector $\mathbf{z}$

\STATE Map the angles $\theta_n$ of each elements of the complex vector $\mathbf{z}$ from $\left[-\frac{\pi}{2},\frac{3\pi}{2}\right)$ to  $\left[-\frac{\pi}{2}, \frac{\pi}{2}\right)$, denoted by $\hat{\theta}_{n}$ as in (\ref{angles})
\STATE Sort $\hat{\theta}_n$ in a non-increasing order and denote as $\hat{\theta}_{m(n)}$

\STATE Construct $\tilde {\mathbf{U}}$ and obtain the set of $N+1$ spreading codes ${\mathcal{U}}$ according to (\ref{setU}), (\ref{tildeU}) and (\ref{U})
\STATE $\mathbf{\bar w}_{opt} = {\rm \arg}\ {\rm \mathop{max}\limits_{\mathbf{\bar w}\in \mathcal{U}}} \ \left|\mathbf{\bar w}^T\mathbf{z}\right|$, obtain $\mathbf{\bar w}_{opt}$ by searching over the set $\mathcal{U}$, and $\mathbf{ w}_{opt} = (\mathbf{ \bar{w}}_{opt}^{(1:N)}./{ \bar{w}}_{opt}^{(N+1)})$
\RETURN $\mathbf{W}(\boldsymbol{ \theta })={\rm diag}(\mathbf{w}_{opt}^H)$
\end{algorithmic} 
\end{algorithm}

\section{Simulation Results}\label{Section4}
We compare and analyze the performance of the proposed DaS method with several other optimization methods, including SDR-SDP, Manopt, and APX~\cite{zhang2022configuring} in the section.

SDR-SDP and Manopt are commonly used for continuous beamforming in previous works, here our numerical experiments show that they work for discrete beamforming as well. 
For SDR-SDP, we apply a standard CVX solver in Matlab~\cite{grant2014cvx} to solve the convex optimization problem obtained from relaxation. 

Exhaustive search always finds the global optimal phase shifts matrix, but is considered only for small-sized RIS due to its high time complexity. 
APX is a discrete optimization method recently proposed in \cite{zhang2022configuring}, mainly for solving binary phase beamforming problems in point-to-point communication systems with linear time complexity.

\subsection{Signal power gains}

\begin{figure*}[!t]
\centering
\subfloat[]{\includegraphics[width=2.35in]{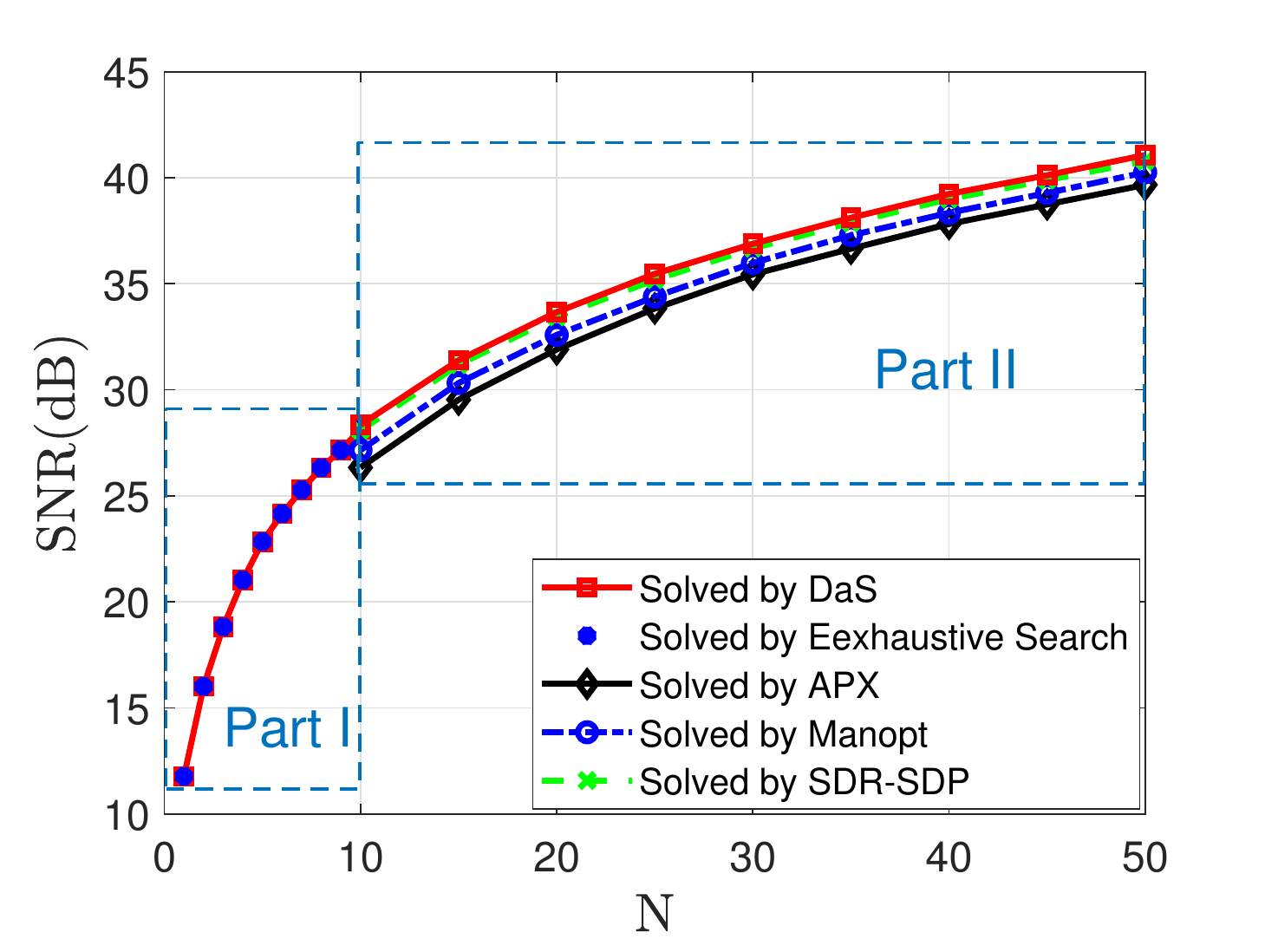}%
\label{fig: Compa}}
\hfil
\subfloat[]{\includegraphics[width=2.35in]{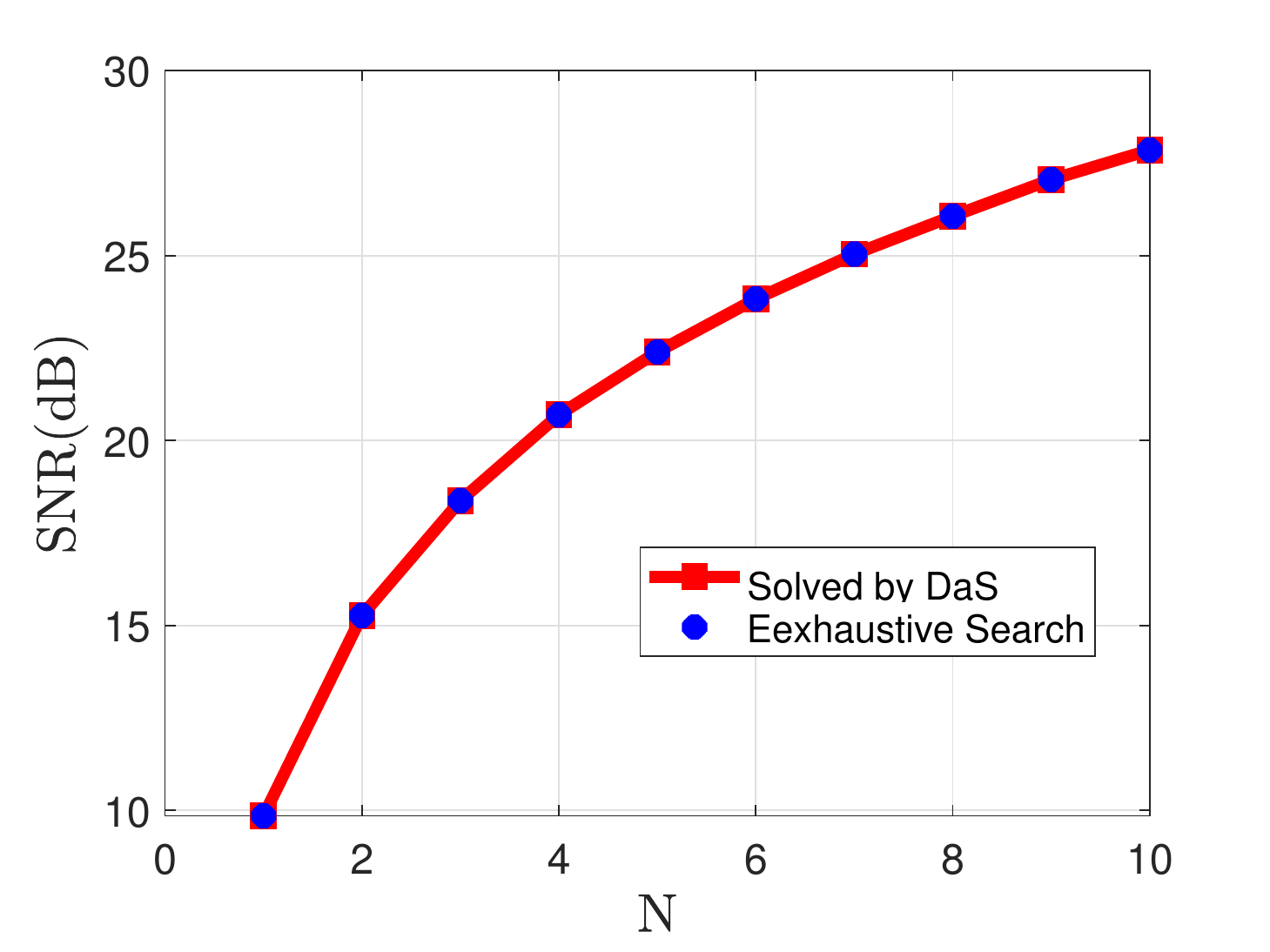}%
\label{fig: CompaExs}}
\hfil
\subfloat[]{\includegraphics[width=2.35in]{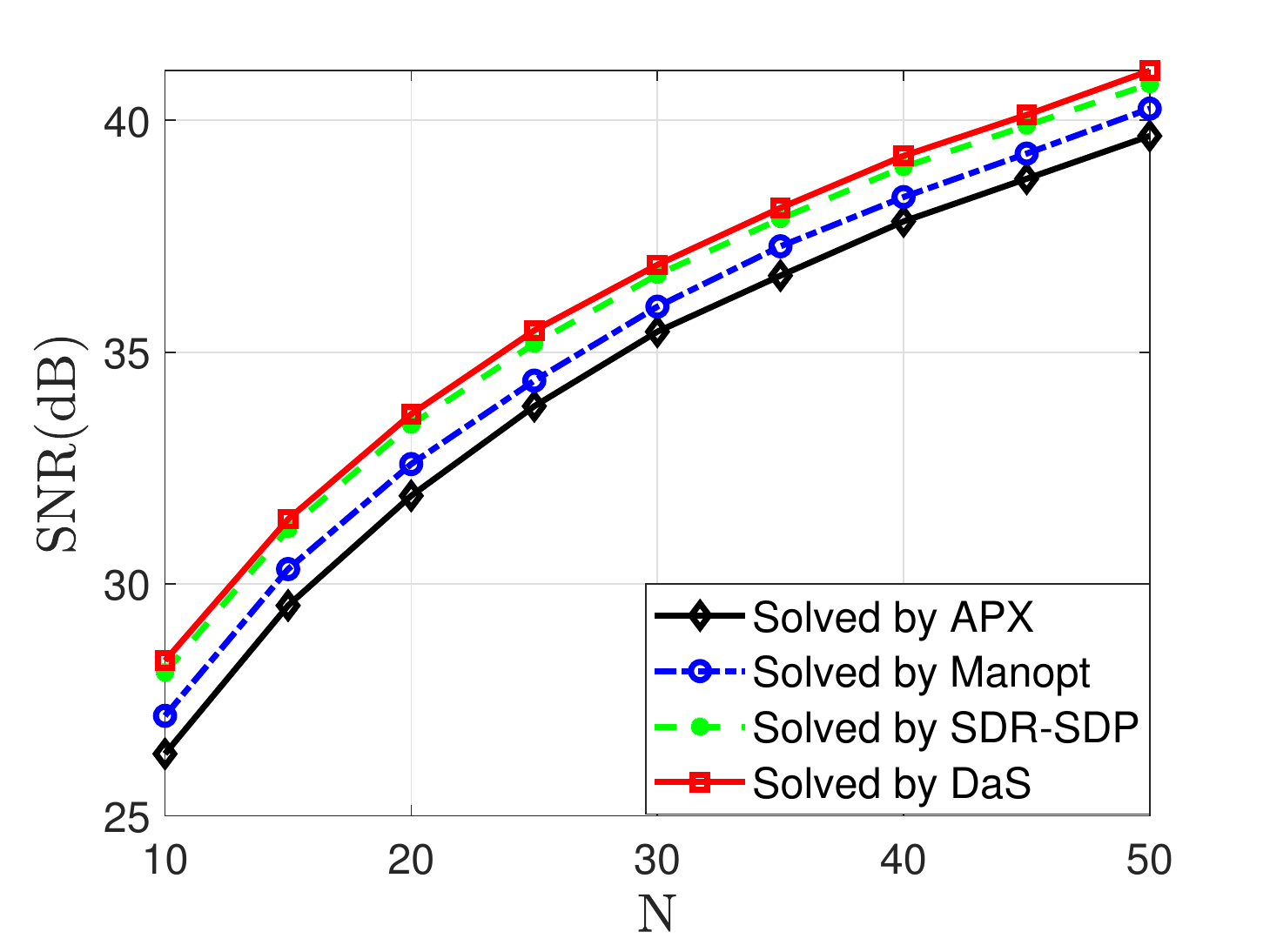}%
\label{fig: Compa3Method}}
\centering 
\caption{Comparison of SNR performance for different optimization methods including SDR-SDP\cite{cui2019secure,elmossallamy2021ris,zhou2020robust}, Manopt \cite{yu2019miso, elmossallamy2021ris,hu2022constant}, and APX \cite{zhang2022configuring} as a function of the number of RIS elements $N$, simulation average over 1000 times. (a) Different methods. (b) Zoom in on Part {\uppercase\expandafter{\romannumeral1}} (with exhaustive search). (c) Zoom in on Part \uppercase\expandafter{\romannumeral2}.}
\label{Fig1}
\end{figure*}

We compare the performance of these methods in terms of their SNR in Fig.\ref{Fig1}. For better visualisation, we zoom in for $N \in [1,10]$ and $N \in [10,50]$ in Fig.\ref{fig: Compa} to get Fig.\ref{fig: CompaExs} and Fig.\ref{fig: Compa3Method}, respectively.  
Fig.\ref{fig: CompaExs} shows that the proposed discrete optimization method in Algorithm~\ref{alg2} can achieve the same SNR performance as the exhaustive search method, which, due to its high time complexity, can only find the global optimal phase shifts for small-sized RIS (say with $N \leq 10$).

Fig.\ref{fig: Compa3Method} zooms in on the Part \uppercase\expandafter{\romannumeral2} of Fig.\ref{fig: Compa}, and compares the SNR performance of the proposed DaS algorithm with existing methods including SDR-SDP, Manopt, and APX. 
These results identify an averaged $0.5$dB, $1$dB, and $2$dB SNR gain of the proposed DaS approach over SDR-SDP, Manopt, and APX, respectively. 
And all methods perform better as the RIS element $N$ increases.

\begin{figure}[!t]
\centering
\includegraphics[width=2.5in]{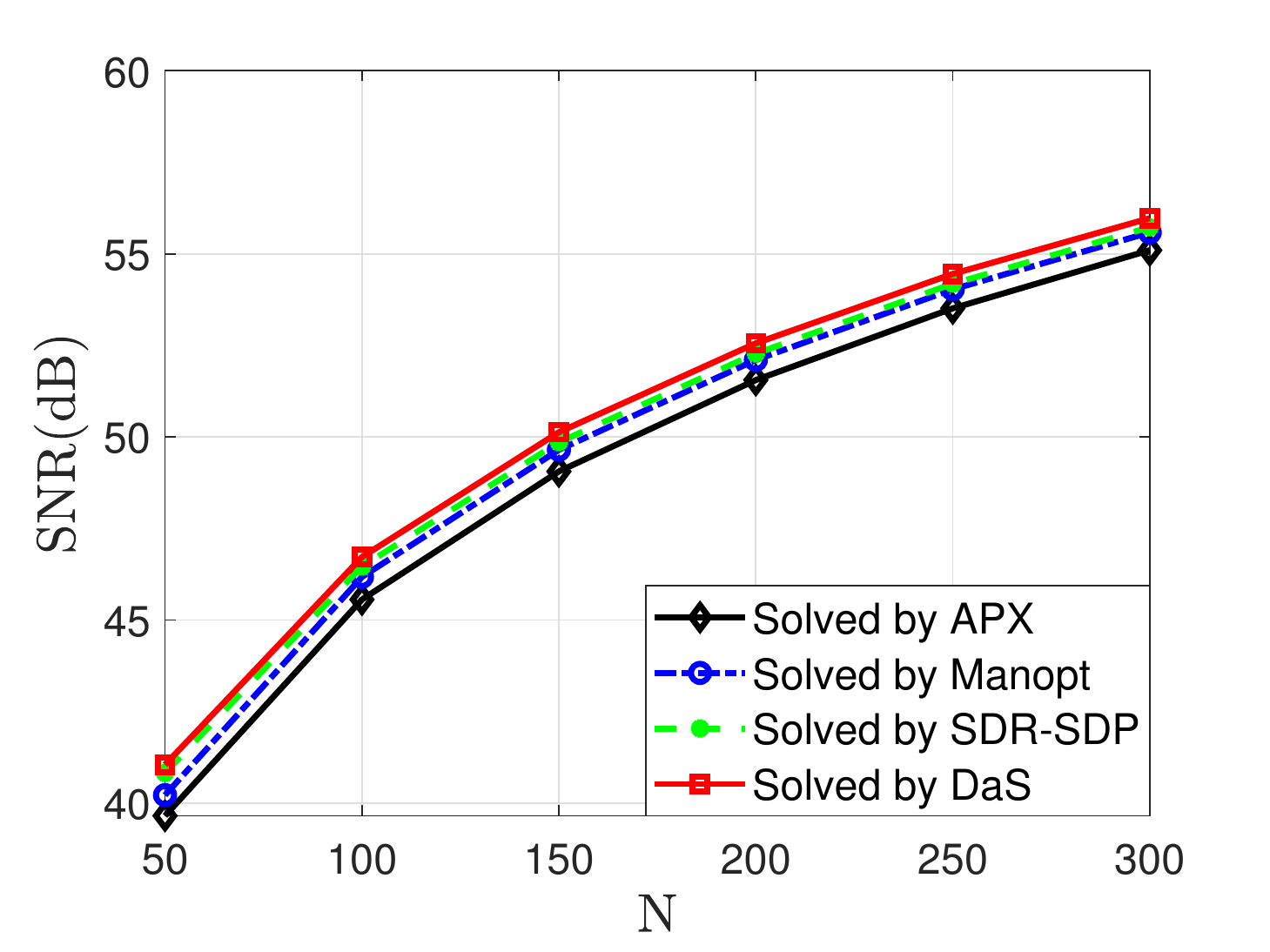}%
\label{fig: Compa2}
\caption{SNR performance for different optimization methods as a function of the number of RIS elements $N$.}
\label{Com4}
\end{figure}

Fig.\ref{Com4} shows that, as $N$ increases, the gaps in SNR between different methods become smaller, but the SNR obtained by the proposed DaS algorithm always differs from that by APX by a margin of approximately 1dB, which is rather significant. 
Meanwhile, the running time of DaS is even (slightly) less, which will be mentioned in the next subsection.

\begin{table*}[!t]
\caption{Total running time with realization $100$ times, of the different optimization methods vs. $N$}
\centering
\begin{tabular}{c|c|c|c|c|c|c|c}
Methods &  $N = 10$ & $N = 50$ & $N = 100$ &$N = 200$ &$N = 500$ & $N=1000$ \\\hline
Exhaustive Search& $0.96$s& {\sf NaN}  & {\sf NaN}   &{\sf NaN}       &{\sf NaN}    &{\sf NaN}   \\
SDR-SDP  &$169.26$s &$191.67$s &$312.45$s & $903.77$s &$10785.55$s & {\sf NaN}                          \\
Manopt & $3.26$s         & $26.11$s &  $48.50$s & $80.95$s &$260.90$s & $950.24$s            \\
APX              &$0.26$s   &$0.39$s   &$0.55$s   &$1.33$s   & $8.67$s &$38.16$s \\
Proposed DaS& $0.23$s       & $0.25$s  &  $0.31$s  & $0.61$s  &$3.80$s   & $21.15$s  
\label{tab: time}

\end{tabular}
\end{table*}


\subsection{time complexity}

\begin{figure}
    \centering
    \includegraphics[width=0.35\textwidth]{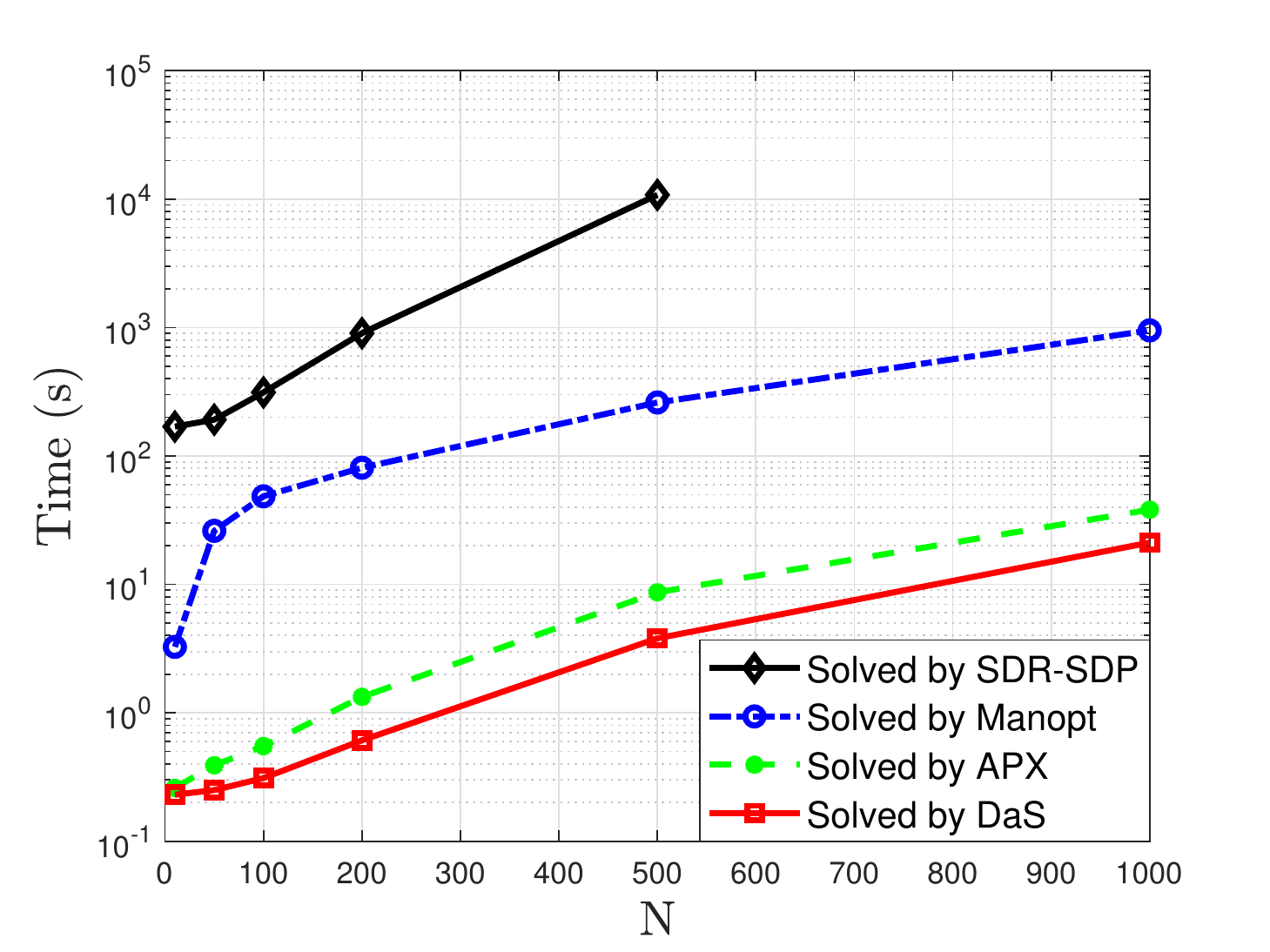}
    \caption{\label{fig: time}Total running time with realization $100$ times, of the different optimization methods vs. $N$  }
\end{figure}

The running time by each of the optimization methods mentioned is shown in Table \ref{tab: time} and is visually displayed in Fig.\ref{fig: time}. 
We observe that the proposed DaS algorithm spends much less time than the other algorithms including SDR-SDP, Manopt, and APX. 
It is worth noting that the proposed algorithm can solve the optimization problem quickly even when $N$ exceeds $1000$, in which case the CVX tools \emph{fail} to provide a solution after running for one hour. 
Such situations are denoted as {\sf NaN} in the paper.
Actually, the DaS approach always obtains the optimal solution in every realization, which means it needs \textbf{just} $\mathbf{0.2115s}$ to update RIS states when the elements number $N = 1000$. Common RIS consists of about $N=100$ elements cost \textbf{only} $\mathbf{0.0031s}$ to finish the update. 

\section{Conclusion and future perspective}\label{Section5}
NP-hard problems are often encountered when using RIS for, i.e., signal enhancement. Existing continuous optimization methods give friendly solutions but converge slowly and produce quantization errors. Discrete methods like exhaustive search have time complexity $\mathcal{O}(2^{N})$, and other discrete methods are at risk of obtaining a suboptimal solution.
In this paper, we propose an optimal design algorithm DaS for $1$-bit RIS-aided wireless communication networks. With this approach, the optimal global solution can be obtained with complexity  $\mathcal{O}(N\log{N})$, which achieved the same SNR performance as the exhaustive search. 
Numerical results show optimal phase shift matrix can be obtained within $0.22$s even $N$ is $1000$, and gains at least $1$dB in SNR than up-to-date APX. 

The proposed Das algorithm can be implemented in high-speed networks to provide service due to its simplicity and low time complexity. And it will be extended to more complex scenarios such as multi-user networks in our future work.


\bibliographystyle{IEEEtran}
\bibliography{Reference}
\end{document}